\documentclass{optica-article}

\journal{opticajournal} 

\articletype{Research Article}
\usepackage{graphicx}
\usepackage{svg}
\usepackage{lineno}
\usepackage{siunitx}
\usepackage{amsmath}
\usepackage{hyperref}

\begin{document}

\title{Robust single-shot 3D fluorescence imaging in scattering media with a simulator-trained neural network}

\author{Jeffrey Alido,\authormark{1} Joseph Greene,\authormark{1} Yujia Xue,\authormark{1} Guorong Hu,\authormark{1}, Yunzhe Li,\authormark{1 \textdagger} Mitchell Gilmore,\authormark{1} Kevin J. Monk, \authormark{2} Brett T. DiBenedictis,\authormark{3} Ian G. Davison,\authormark{2} Lei Tian,\authormark{1,3,*}}

\address{\authormark{1}Department of Electrical and Computer Engineering, Boston University, Boston, MA, 02215, USA\\
\authormark{2}Department of Biology, Boston University, Boston, MA 02215, USA\\
\authormark{3}Department of Psychology and Brain Sciences, Boston University, Boston, MA 02215, USA\\
\authormark{4}Department of Biomedical Engineering, Boston University, Boston, MA, 02215, USA\\
\authormark{\textdagger}Current address: Department of Electrical Engineering and Computer Sciences, University of California, Berkeley, California, 94720, USA}

\email{\authormark{*}leitian@bu.edu} 


\begin{abstract*} 
Imaging through scattering is a pervasive and difficult problem in many biological applications. The high background and the exponentially attenuated target signals due to scattering fundamentally limits the imaging depth of fluorescence microscopy. Light-field systems are favorable for high-speed volumetric imaging, but the 2D-to-3D reconstruction is fundamentally ill-posed, and scattering exacerbates the condition of the inverse problem. Here, we develop a scattering simulator that models low-contrast target signals buried in heterogeneous strong background. We then train a deep neural network solely on synthetic data to descatter and reconstruct a 3D volume from a single-shot light-field measurement with low signal-to-background ratio (SBR). We apply this network to our previously developed Computational Miniature Mesoscope and demonstrate the robustness of our deep learning algorithm on scattering phantoms with different scattering conditions. The network can robustly reconstruct emitters in 3D with a 2D measurement of SBR as low as 1.05 and as deep as a scattering length. We analyze fundamental tradeoffs based on network design factors and out-of-distribution data that affect the deep learning model’s generalizability to real experimental data. Broadly, we believe that our simulator-based deep learning approach can be applied to a wide range of imaging through scattering techniques where experimental paired training data is lacking.
\end{abstract*}

\section{Introduction}
Fluorescence imaging is crucial in biological research since it permits observation of gene expression and molecular interactions in cells and tissues\cite{mertz_strategies_2019,weisenburger_guide_2018}. One-photon systems, such as widefield microscopes, are often employed to measure fluorescence, but the imaging depth is limited by tissue scattering which degrades and obscures target signals\cite{cheng_development_2019,horton_vivo_2013}. This, along with non-specific fluorescence that contributes to a high level of hazy, spatially heterogeneous background, resulting in measurements with a low signal-to-background ratio (SBR), where objects embedded in deeper layers are nearly imperceptible to both humans and machines. 

Light scattering in one-photon fluorescence imaging systems corrupts measurements in an almost non-invertible way, of which the most significant aspect impairing image formation is the attenuation of ballistic photons and increased background contributions\cite{xue_single-shot_2020,skocek_high-speed_2018}. A conventional approach to recover signals from scattering-contaminated fluorescence measurements is to perform background removal using standard image processing algorithms\cite{xue_single-shot_2020,kauvar_cortical_2020}. The drawback of this approach is that low-contrast target signals are often removed along with the background since the algorithm cannot discriminate low-SBR target signals from the background. As a result, it is difficult to reconstruct objects beneath shallow layers of scattering media as their signals carried by the ballistic photons attenuate exponentially and are further obscured by strong background. While this results in measurements with low SBR, the object information is not completely lost\cite{moretti_readout_2020}. The challenge becomes how to robustly utilize the information encoded in the scattered light in the presence of strong background. 

\begin{figure}[t!]
    \centering
\includegraphics[width=1\linewidth]{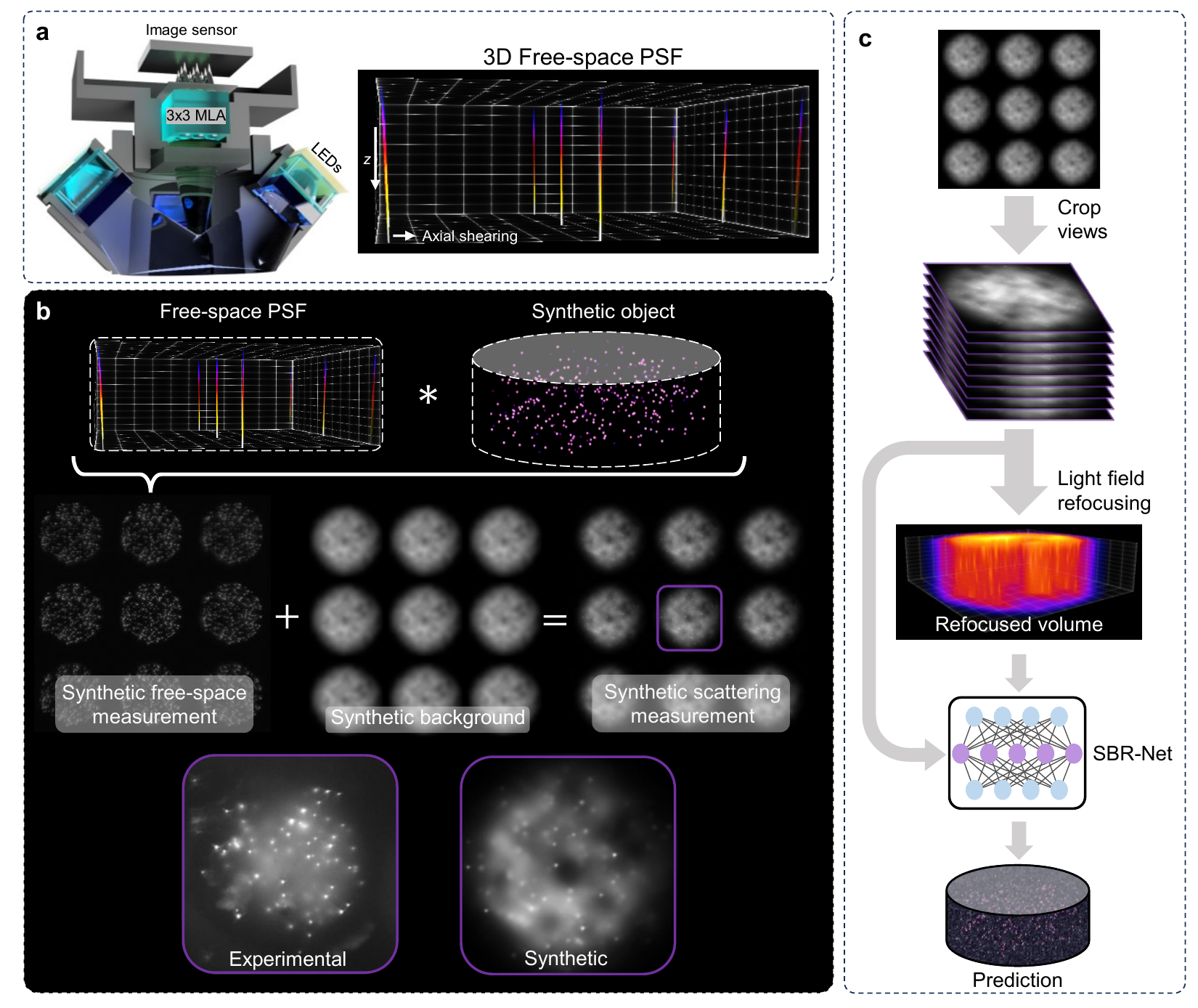}
  
  \caption{\textbf{Overview of our framework.} (a) CM$^2$ captures 3$\times$3 views in each 2D measurement. (b) The forward model for generating synthetic data. The final synthetic scattering measurement is the sum of synthetic free-space measurement and synthetic background. A visualization comparing simulated and experimental data is shown. (c) SBR-Net reconstruction fuses LF refocusing enhancement and parallax information from the 3$\times$3 views to reconstruct objects from low-SBR measurements.}
  \label{fig: fig1 overview}
\end{figure}

Recently, deep learning (DL) methods have been applied to imaging through scattering\cite{li_deep-3d_2022,li_deep_2018,tahir_adaptive_2022,wijethilake_deep2_2022,zhang_rapid_2023} due to its ability to solve highly ill-posed inverse problems. An outstanding challenge to apply DL techniques in real experiments is how to collect training data. One strategy is to collect experimentally paired data to learn the inverse mapping to descatter images and reconstruct high contrast objects\cite{li_deep-3d_2022,li_deep_2018}; however, obtaining ground-truth reference is costly and time-consuming. Another strategy is to employ a physics-based imaging model to aid in learning the inverse mapping by either integrating it directly in a self-supervised learning scheme\cite{liu_recovery_2022} or by generating synthetic paired training data offline for supervised learning\cite{tahir_adaptive_2022,wijethilake_deep2_2022,zhang_rapid_2023}. While some of these methods have shown state-of-the-art results, they are limited to reconstructing objects under the specific scattering and imaging conditions on which they were trained. Therefore, these methods are less robust to out-of-distribution (OOD) data caused by experimental deviations that are common in biomedical imaging applications, such as background fluorescence and low SBR.

High-speed volumetric imaging is crucial in biomedical research\cite{mertz_strategies_2019}. Light field (LF) microscopy is a computational imaging technique that enables single-shot 3D fluorescence imaging\cite{xue_single-shot_2020,xue_deep-learning-augmented_2022,yanny_miniscope3d_2020,skocek_high-speed_2018,guo_fourier_2019}. However, the ill-posedness of the 2D-to-3D LF inverse problem is exacerbated by scattering. Nonetheless, LF measurements contain useful information within the tomographic redundancies that can aid with imaging through scattering\cite{nobauer_video_2017,pegard_compressive_2016}. For example,  angular information in LF measurements augmented with temporal fluctuation information of fluorescent objects allow devising 3D reconstruction algorithms robust to volumetric scattering\cite{nobauer_video_2017,pegard_compressive_2016}. A recent study developed an incoherent multi-layer model for scattering-robust 3D fluorescence reconstruction without temporal information\cite{zhang_computational_2021}. However, these methods generally suffer from high model complexity that renders the inversion to be highly time intensive.

Here, we present a robust supervised DL approach for single-shot 3D fluorescence imaging through scattering. We demonstrate our method on our previously developed LF-based Computational Miniature Mesoscope (CM$^{2}$)\cite{xue_single-shot_2020,xue_deep-learning-augmented_2022} (Fig. \ref{fig: fig1 overview}a). 
To overcome the challenge of generating paired training data, we devise a synthetic data generation process that integrates the scattering-free imaging model of CM$^{2}$ and a synthetic scattering-induced background model. Importantly, the synthetic background model is computationally simple and highly generalizable to different scattering conditions. The added background contains random slow variations to account for inhomogeneities due to tissue scattering that are modeled by low-pass filtered value noise\cite{noauthor_value_2021}, and the background strengths are varied to model different SBR levels (Fig. \ref{fig: fig1 overview}b). To utilize the angular information contained in the LF measurement for 3D reconstructions, we develop SBR-Net that combines the functionalities of view-synthesis and enhancement of LF refocusing with a dual-branch structure\cite{xue_deep-learning-augmented_2022}, augmented with a variance stabilization strategy  (Fig.~\ref{fig: fig1 overview}c). 
Our training process of SBR-Net is designed to generalize to different scattering conditions. We generate paired training data by simulating diverse and low SBR LF measurements with SBRs ranging from 1.1 to 3.0, and their corresponding ground-truth volumes. The network takes input of the low-SBR data and reconstructs 3D fluorescent emitters embedded in scattering media. Our simulation results and quantitative analysis show that SBR-Net can accurately reconstruct emitters located at depths up to one scattering length deep inside the scattering media across a broad range of SBR and scattering conditions. 

A major achievement we demonstrate is that SBR-Net trained solely on synthetic data generalizes well to real experiments. Using CM$^{2}$ as testbed paired with the simulator-trained SBR-Net, we demonstrate single-shot 3D reconstruction from 2D measurements of fluorescent objects having SBRs as low as 1.05. We demonstrate our algorithm on three controlled scattering phantoms of scattering lengths of 279, 182, and 72$\mu$m with added fluorescent background where the target object is a mixture of 10 and 15$\mu$m fluorescent beads. We also piloted a study on a fixed mouse brain slice. We benchmark the results with confocal microscopy measurements. SBR-Net recovers emitters embedded as deep as one scattering length in all scattering phantoms with 25$\mu$m axial resolution and shows promising results on complex brain tissues. 

In addition, we provide insights into how to design robust DL models for scattering inverse problems. Our detection-based analysis of our algorithm reveals how the SBR range in simulated training data affects the generalization to real experimental measurements, where the choice of SBR range in the training data has a tradeoff between the imaging depth penetration and false positives, equivalently, the robustness-accuracy tradeoff. We also analyze the bias-variance tradeoff based on the number of unique training data pairs to understand the considerations for network generalization performance. 

Broadly, we believe that our simulator-based DL approach for improving reconstruction from low contrast measurements may open up avenues to improve computational strategies for imaging through scattering in many applications, such as neural imaging and intravital imaging.

\section{Methods}
\subsection{CM\textsuperscript{2} imaging system}
We use our previously developed Computational Miniature Mesoscope (CM$^2$) \cite{xue_single-shot_2020,xue_deep-learning-augmented_2022} to demonstrate our descattering algorithm. Briefly, CM$^2$ uses a 3$\times$3 microlens array (MLA) as the sole imaging element to collect 9 LF views of the object volume. The numerical aperture (NA) of each lenslet is ~0.05, and the imaging system has a magnification of ~0.52. We use a Sony IMX-178 image sensor with a pixel size of 2.4$\mu$m, rendering the object space sampling to be 4.15$\mu$m. CM$^2$ is equipped with an absorption and excitation filter that rejects light outside the emission band of green fluorescence and blue light, respectively. Further specifications and assembly instructions may be found on our open-source GitHub\cite{noauthor_httpsgithubcombu-cislcomputational-miniature-mesoscope-cm2_nodate}.

Within a lateral field of view (FOV) of 2mm in diameter, CM$^2$’s system response is slice-wise shift-invariant\cite{xue_single-shot_2020}, so we constrain our objects of interest to be at most 2mm in diameter. Using a motorized $z$-stage and a 5$\mu$m pinhole source, we collect a defocused stack of on-axis point spread functions (PSF) from $z$ = -225$\mu$m to $z$ = 375$\mu$m with a 25$\mu$m increment to characterize the system's free-space PSF over an axial depth range of 600$\mu$m. All measurements are taken with an exposure time of 50ms and a digital gain of 1$\times$.

\subsection{Forward model}
Our scattering model is comprised of a free space, laterally shift-invariant model combined with added synthetic background to produce a scattering measurement $g$ as
\begin{equation}
    g(u,v)=\alpha \sum_{z} \mathrm{PSF}(u,v;z) \circledast  V(u,v;z) + \mathrm{BG},
    \label{eq: eq1 convolution fwdmdl}
\end{equation}
where $\circledast$ denotes a 2D convolution, $(u,v)$ are the pixel coordinates, $\mathrm{PSF}(u,v;z)$ is the measured free-space PSF at the axial plane $z$, $V(u,v;z)$ is the synthetic ground-truth fluorescent volume, $\mathrm{BG}$ is the synthetic background, and $\alpha$ is a scalar to control the SBR. 

The BG term is drawn from Gaussian-blurred random value noise. The blur kernel size is uniformly distributed between 31.2 and 48$\mu$m, which accounts for background statistics observed in experimental phantoms across different bead densities and scattering conditions. Value noise is a computer graphics tool to generate procedural textures by interpolating between integers on a random lattice\cite{noauthor_value_2021}. Value noise is similar to Perlin noise, which has been used for modeling background fluorescence\cite{mockl_accurate_2020}. We use open source code\cite{262588213843476_2d_nodate} to generate raw value noise of size 600$\times$600 pixels,  corresponding to a 1.4mm $\times$ 1.4mm FOV. A sample of raw value noise and its low-pass filtered version are shown in Fig. S1. 
The last step to generate the BG term is multiplying the Gaussian blurred result with a circular Gaussian envelope mask to account for the circular FOV and the apodization of the imaging optics. Each measurement contains 2076$\times$3088 pixels across an 8.6$\times$12.8mm FOV.

We note that the NA of each lenslet in CM$^2$ is 0.05, which is low enough that we may neglect scattering-induced width broadening of the PSF\cite{cheng_development_2019}. For scattering-induced width-broadening to become significant enough in our imaging system, the depth range would have to be around 5 scattering lengths, which is beyond our application with one-photon imaging.

\subsection{Synthetic training and testing data generation}
All data synthesis and analysis were carried out in MATLAB R2019b. 
We generate $V(u,v;z)$ by simulating spheres with diameters normally distributed around a mean of 15$\mu$m with standard deviation (std) of 2$\mu$m, and brightness normally distributed around a mean of 0.8 with std of 0.1. The spheres were first generated on a 5$\times$ finer grid with voxel size of 0.83$\mu$m $\times$ 0.83$\mu$m $\times$ 5$\mu$m, then 5$\times$5$\times$5 average binned to make the ground-truth volume have the same discretization as our reconstruction. We generate our synthetic volumes with emitter densities normally distributed with mean 180 emitters/mm$^3$ and std of 118 emitters/mm$^3$.

To control the SBR, we first normalize the free-space measurement term and the background term in Eq.~\eqref{eq: eq1 convolution fwdmdl} to be between 0 and 1 before combining them. We compute the peak intensities of all emitters in the free space measurement using $\texttt{imregionalmax}$ function and average the values to get the average target signal $\overline{S}$. We compute the mean background signal $\overline{BG}$ by averaging all the pixel values in the value noise sample. Defining SBR as
\begin{equation}
    \mathrm{SBR} = \frac{\alpha \overline{S} + \overline{\mathrm{BG}}}{\overline{\mathrm{BG}}}.
    \label{eq: sbr eq}
\end{equation}
we then calculate the value of $\alpha$ for a desired SBR of a scattering measurement to use in Eq.~\eqref{eq: eq1 convolution fwdmdl}.
 
The final scattering measurement $f$, includes signal-dependent mixed Poisson-Gaussian (MPG) noise \cite{foi_practical_2008} with noise parameters calibrated from experimental data using the following equation
\begin{equation}
    f = g + \sigma(g)\cdot\xi,
    \label{eq: f = g sigmag}
\end{equation}
where $\xi$ is a random variable drawn from a normal distribution, and $\sigma(g)$ is the signal-dependent std given by $\sigma(g) = \sqrt{ag+b}$,
where $a$ and $b$ are parameters for Poisson and Gaussian noise, respectively. The BG term was considered as signal for computing signal-dependent noise. Calibration determined these parameters to be $a$ = 1.49e-4 $\pm$ 0.57e-4 and $b$ = 5.41e-6 $\pm$ 2.78e-6. MPG noise was added during network training in an online fashion where parameters $a$ and $b$ are randomized, and the measured uncertainties are the std for a normal distribution.

To generate free-space synthetic training data, we use Eq. \eqref{eq: eq1 convolution fwdmdl} with BG = 0 and $\alpha = 1$, and the same steps are taken for adding noise. For generating training data for a benchmark BGR-Net (detailed later), we first add MPG noise to the scattering measurement offline, and then perform background removal. No further noise was added online during training.

To generate the testing data, we add an additional signal attenuation model based on Beer-Lambert's law, neglecting absorption, given by $I(z) = I_0\exp(-z/\ell_s)$, where $\ell_s$ is the scattering mean free path. As a note, we do not include this attenuation for the training data so that we may have better control over the SBR of simulated measurements.

\subsection{Network implementation}

Our network design follows a 2D ResNet structure, where the depth dimension is the channel dimension of the tensors. The network design has two branches that take as input different but equivalent forms of the measurement, and then fused together in a final layer. 
Each branch is comprised of 20 ResBlocks that contain, in order: a 3$\times$3 convolutional layer, batch normalization (BN), ReLU, another 3$\times$3 convolutional layer, and BN. 
The input to each ResBlock is added to the output of the same ResBlock. The inputs for the two branches are the stack of 9 views, and the LF refocused volume (details in Sec.~S12), whose channel dimension are expanded from 9 to 48, and 24 to 48, respectively, in an initial expansion 3$\times$3 convolutional layer. 
The input to the first ResBlock is added to the output of the 20th ResBlock. After 20 ResBlocks, the branches are fused together and a final 3$\times$3 convolutional layer squeezes the 48 channels to 24, which are the number of axial slices in the ground-truth volume. 
An illustration of the architecture is shown in Fig. S5. There are in total $N = 42$ 3$\times$3 convolutional layers along the forward path of the network, resulting in a receptive field of 83 pixels or 344.45$\mu$m.  We also demonstrate a variance stabilization strategy to substantially improve ResNet-based training loss convergence, shown in Fig. S15, which we discuss in detail in Sec. S9.

We use cross-entropy as the loss function to promote sparsity in the reconstruction\cite{li_deep_2018}:
\begin{equation}
    \mathrm{CE}(y,\hat{y}) = \sum_i u_i\log\hat{y}_i + (1-y_i)\log(1-\hat{y}_i),
    \label{eq: cross entropy}
\end{equation}
where $y_i$ and $\hat{y}_i$ are  the ground truth and reconstruction intensity, respectively, at voxel $i$.

\subsection{Training details and benchmark networks}

We generate 500 pairs of paired scattering training data for training SBR-Net.
In addition, to compare with more traditional approaches, we train a free-space network FS-Net on 500 pairs of free space data. The input of FS-Net is the scattering measurement processed with a traditional background removal algorithm (details in Sec.~S11). 
Finally, we also take the 500 scattering measurements and perform background removal and then train a background-removed network BGR-Net.  All three networks have the same architecture and training parameters.

The 500 pairs were split 80/20 for training and validation. 
Training was carried out on a single NVIDIA V100 GPU with 16 GB memory using the PyTorch 1.9.0 framework. We initialize the weights using He initialization, which is optimized for ReLU non-linearities to prevent exploding or vanishing forward propagation signals\cite{he_delving_2015}. We set the bias for each convolutional layer to be false. The inputs are B $\times$ C $\times$ 224 $\times$ 224 patches from the original B $\times$ C $\times$ 512 $\times$ 512 data, and the patch locations are randomized online during training. The batch size was heuristically set to be 12 to maximize memory availability on the GPU. The data and the weights of the network are single precision floating point numbers at 16-bit, which we accomplish using PyTorch’s automatic mixed precision (AMP) package. We use Adam for the optimizer and cosine annealing\cite{loshchilov_sgdr_2017} for the training schedule with an initial learning rate of \SI{1e-3} and a period of 30 epochs. We set the maximum training time to be 48 hours and SBR-Net settled on a local minimum that was found after 28 hours, FS-Net 39 hours, and BGR-Net 12.5 hours. 

During inference, the entire C $\times$ 512 $\times$ 512 data is passed through the network. For FS-Net and BGR-Net, we remove the background and then pass the input to the networks. For SBR-Net, we pass the raw, low-SBR scattering measurement input to the network. The inference runtime for one volume is 0.11 seconds on average using an Intel Xeon E5-1620 v4 3.5 GHz CPU and 0.022 seconds on average using a Nvidia Quadro RTX 8000 48GB GPU. 

\section{Results}
\subsection{Statistical comparison of simulated and experimental measurements}
We begin by comparing the statistical properties of experimental and synthetic data to validate our scattering and background model. As shown in Figs. S2 - S4, the synthetic measurement shares visual qualities with experimental measurements. Additionally, our background model provides good statistical matching with experiment, given by comparing their spatial frequencies, power spectra, and intensity histogram, which is shown to be an important reason for the robustness of neural networks\cite{deng_learning_2020}. We also perform principal component analysis (PCA) for each measurement domain, shown in Fig. \ref{fig: pca and bgr test}a, and see that the first two principal components of the simulated scattering measurement clusters closely with those of experimental scattering measurements.

\subsection{Simulated fluorescent beads reconstruction}
We first quantify the performance of SBR-Net on synthetic scattering samples under different scattering conditions. Different from the training data, we further incorporate a signal attenuation model ($I(z) = I_0\exp(-z/\ell_s)$) in the testing data to assess the performance of SBR-Net on more realistic and quantifiable scattering scenarios. We experimented on three different synthetic phantoms with scattering lengths of 80, 160, and 320$\mu$m. 

\begin{figure}[t!]
    \centering
    \includegraphics[width=1\linewidth]{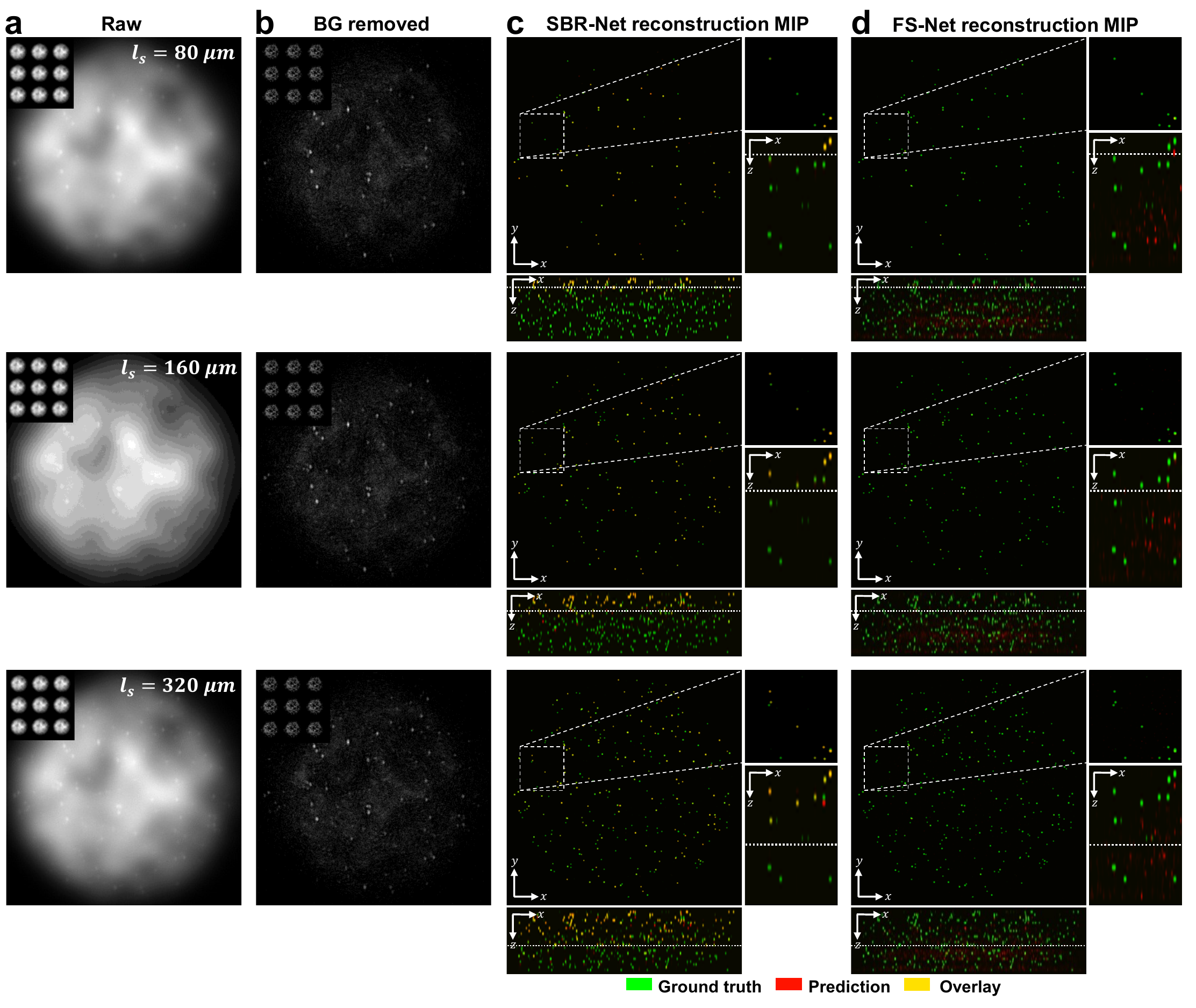}
  \caption{\textbf{Qualitative reconstruction results for synthetic scattering phantoms.} The scattering lengths of test samples grow quadratically, including Row 1: $\ell_s = 80 \mu m$, Row 2: $\ell_s = 160 \mu m$, and Row 3: $\ell_s = 320 \mu m$. (a) Raw measurement of the central view with the full 3$\times$3-view measurement as the inset. (b) Background (BG) removed measurement. (c,d) Maximum intensity projections (MIPs) of the reconstructions of (c) SBR-Net and (d) FS-Net. The ground-truth beads are shown in green, predictions in red, and overlay in yellow. XY MIPs are shown for volumes as deep as one scattering length. The dotted line in the XZ MIPs represent a distance of one scattering length, and we visually observe good emitter reconstruction and localization up to one scattering length for SBR-Net, and almost no predicted emitters beyond one scattering length. FS-Net performs poorly even within shallow layers, while also generating false positives.}
  \label{fig:fig2 qualitative}
\end{figure}

\begin{figure}[t!]
    \centering
    \includegraphics[width=1\linewidth]{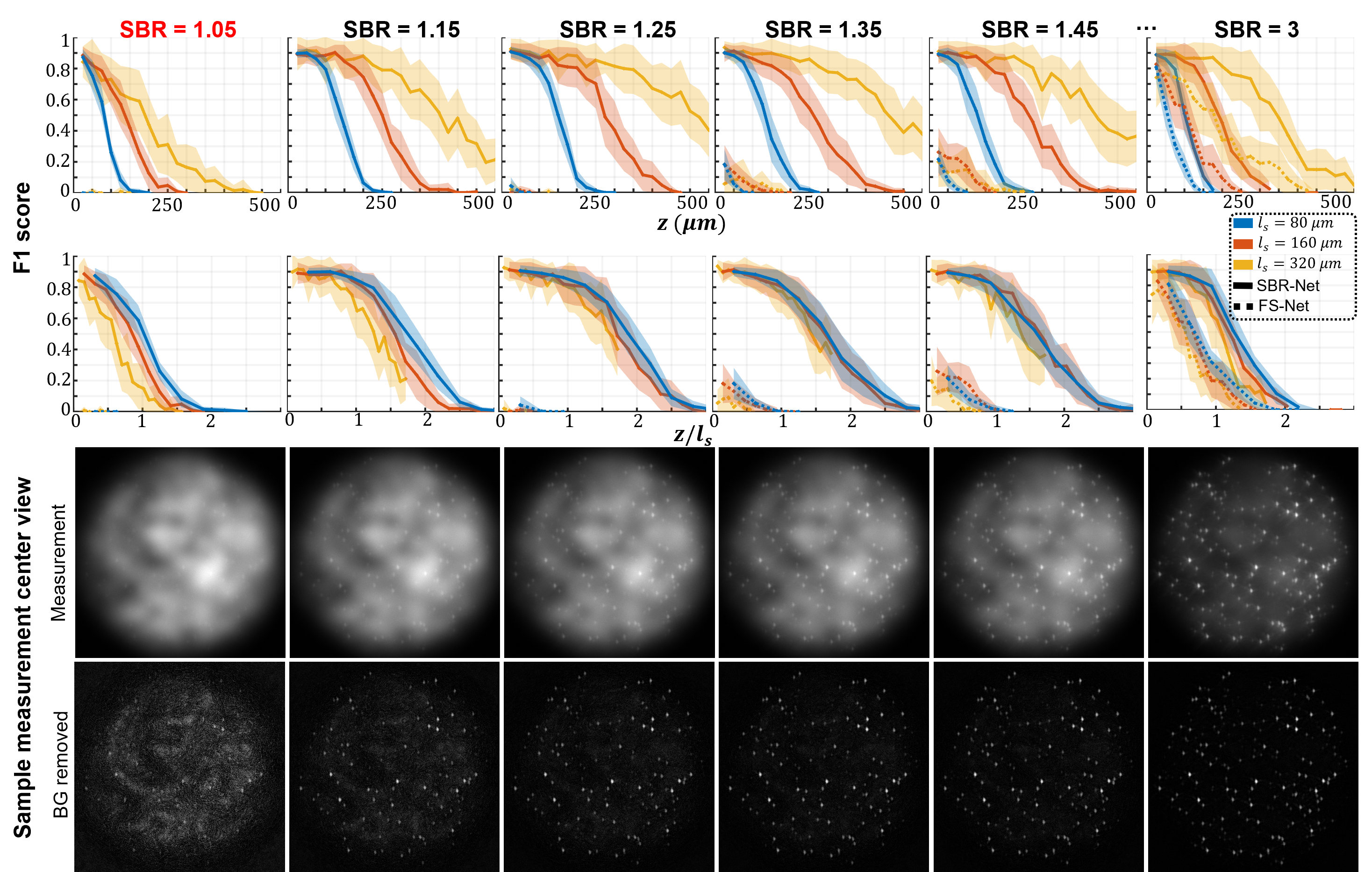}
  
  \caption{\textbf{Simulation test data F1 score and visualization of different SBR measurements.} The localization performance over physical and normalized depth for different scattering length samples is shown for increasing SBR. Each column corresponds to data with the labeled SBR and the images are sample measurements of data with the labeled SBR. The labeled SBR values are the average SBR values for the emitters in the first depth layer before any optical attenuation. Similar to the optical signal, the SBR of an emitter would decay exponentially over increasing depth of the emitter, which is why we observe a systematic decay of performance for all scattering lengths, as seen in the normalized depths curves. The line represents the average over 25 samples and shaded regions represent standard error.}
  \label{fig: fig3 synthetic plots}
\end{figure}

In Fig. 2, we visually assess the SBR-Net reconstructions, which is further quantified using the F1 score (see Sec.~S13) over depths for all SBR from 1.05 to 3.0 in Fig.~\ref{fig: fig3 synthetic plots} and Fig.~S6. When examining the results with physical depth ($z$), we see that the F1 score generally decays with depth. The larger the scattering length ($\ell_s$) the slower the decay rate, indicating a deeper imaging capability compared to samples with shorter scattering lengths. To provide additional physical insights, we further examine the curves with the normalized depth by the scattering length ($z/\ell_s$). For a given SBR, the three curves corresponding to different scattering lengths concentrate around the same trend line. In all SBR cases, SBR-Net is able to reconstruct emitters in 3D with an F1 score of around 0.9 for up to one scattering length. Higher SBR generally results in slightly improved performance at deeper depths. SBR-Net was trained on scattering measurements with an SBR range of 1.1 to 3.0, which explains why there is weaker depth penetration for the OOD worse SBR = 1.05 case. However, we show in our ablation study that it is undesirable to train SBR-Net on lower SBRs due to poorer generalization performance to experimental data. The results for the SBR = 3.0 case are slightly worse than those of lower SBR, which can be explained by the network having seen more examples of images with lower contrast than higher, resulting in poorer reconstructions for higher contrast data.

We benchmark the performance with FS-Net trained on background removed scattering measurements. SBR-Net significantly outperforms FS-Net in terms of reconstruction accuracy and depth penetration, where SBR-Net has a higher F1 score over all depths for all SBR from 1.05 to 3.0. Additionally, SBR-Net has a higher precision and recall score than FS-Net, shown in Fig.~S6. FS-Net performs poorly due to the background removal step removing target emitters along with the background. Emitters embedded more deeply in scattering tissue are attenuated more strongly according to Beer-Lambert’s law and thus have a lower SBR, making its morphological features similar to that of the slowly varying background signal. Additionally, there are severe hallucination artifacts.

\begin{figure}[ht]
    \centering
    \includegraphics[width=1.0\linewidth]{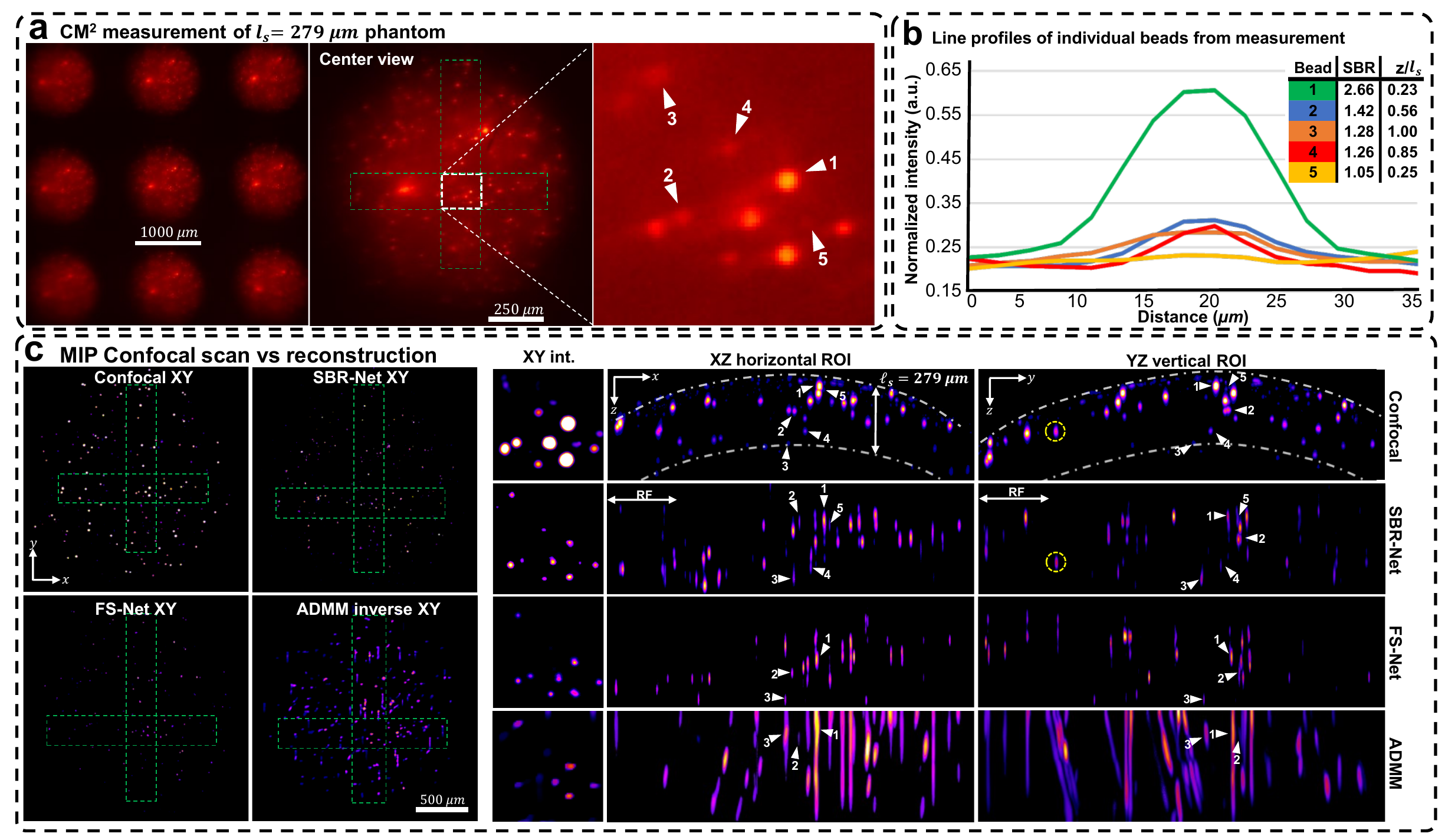}
  \caption{\textbf{Reconstruction results for an} $\mathbf{\ell_s= 279\mu}$\textbf{m scattering phantom.} (a) Raw CM$^2$ measurement with a zoom-in of a small ROI with labeled beads for quantifying reconstruction performance. (b) Line profiles of the labeled beads in the raw measurement. Each bead's depth relative to the surface of the phantom (based on confocal measurements) and their measurement SBRs are listed. (c) MIPs of confocal microscopy 3D measurements, reconstructions with SBR-Net, FS-Net, and model-based algorithm. The dashed-dotted line in the confocal XZ/YZ MIPs represent a distance of one scattering length from the surface of the phantom. SBR-Net can recover all 5 beads at their correct depth location, which is as deep as a complete scattering length. SBR-Net also reconstructs and localizes an emitter with a measurement SBR of 1.05. FS-Net and ADMM fail to reconstruct 2 emitters of the 5 with the lower SBRs and localize the remaining 3 with poor accuracy compared to that of SBR-Net. Additionally, the XY MIPs show that SBR-Net provides background rejection while retaining low SBR emitters, which FS-Net and ADMM remove.}
  \label{fig: fig4 279exp}
\end{figure}

\subsection{Experimental scattering phantoms reconstruction}
We experimentally evaluate SBR-Net on three scattering phantoms of different scattering lengths, including 72, 182, and 279$\mu$m (details in Sec.~S5). In Figs.~4 and 5, we show the CM$^{2}$ measurements and quantify emitters’ SBR and depth in a small region of interest (ROI) for the $\ell_s = 279$ and $72\mu$m phantom, respectively. We also show the XZ and YZ MIPs for ROIs of the reconstruction across the object’s FOV and inspect localization and reconstruction accuracy. We benchmark our results against confocal microscopy measurements (details in Sec.~S14) and our previously developed model-based reconstruction algorithm\cite{xue_single-shot_2020} (details Sec.~S10). The result for the $\ell_s = 182\mu$m phantom is shown in Fig.~S8.

\begin{figure}[t!]
    \centering
    \includegraphics[width=1.0\linewidth]{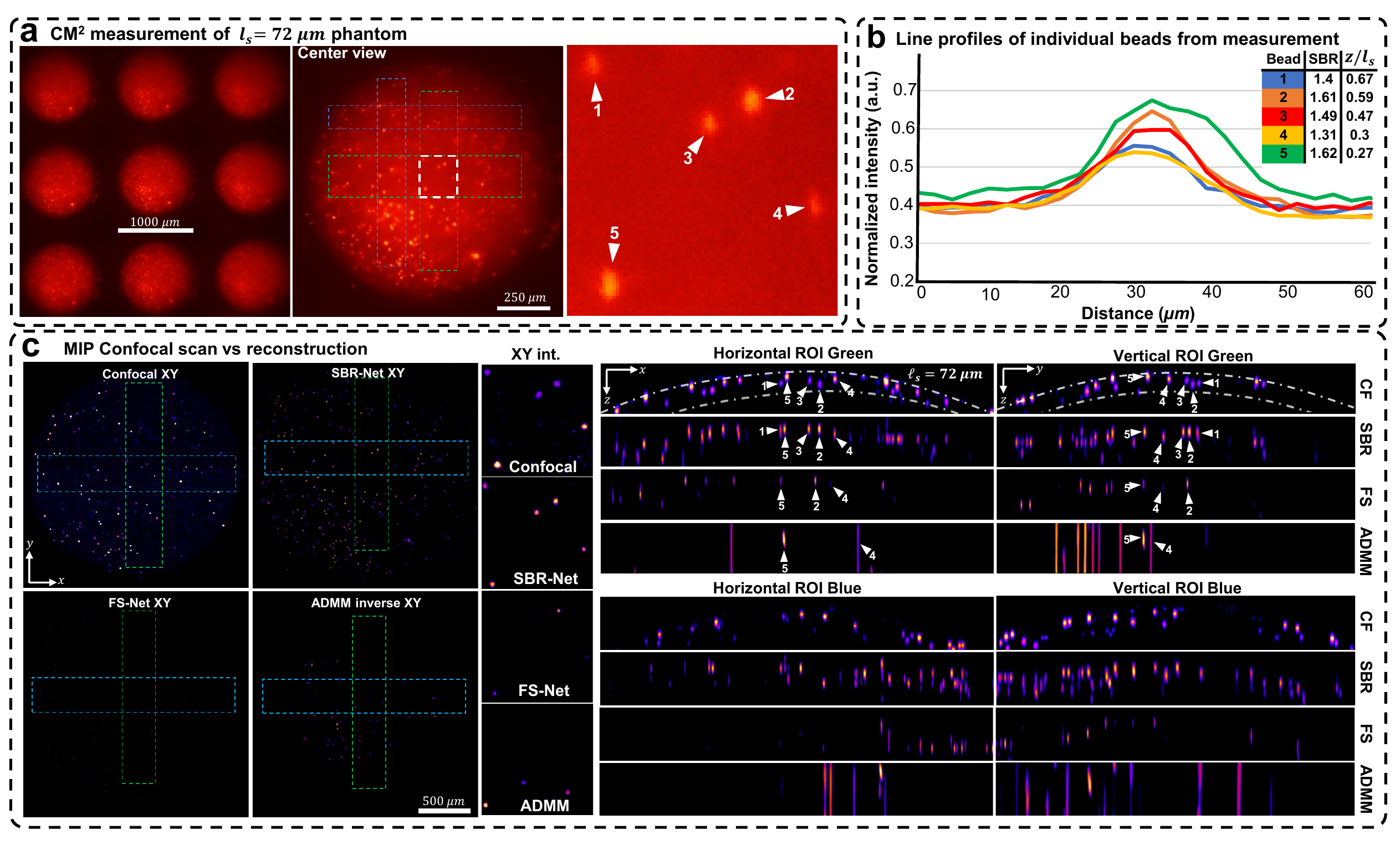}
  \caption{\textbf{Reconstruction results for an} $\mathbf{\ell_s= 72\mu}$\textbf{m phantom.} (a) Raw CM$^2$ measurement with a zoom-in of a small ROI with labeled beads for validating reconstruction performance. (b) Line profiles of the labeled beads in the raw measurement. Each bead's depth relative to the surface of the phantom and their measurement SBRs are listed. (c) MIPs of confocal microscopy 3D measurements, reconstructions with SBR-Net, FS-Net, and model-based algorithm. The dashed-dotted line in the confocal XZ/YZ MIPs represent a distance of one scattering length from the surface of the phantom. SBR-Net is able to recover all 5 beads at their correct depth location. FS-Net and ADMM both fail to reconstruct all 5 particles. Across a large FOV, SBR-Net achieves consistent reconstructions of low SBR particles, while FS-Net and ADMM fail to reconstruct them.}
  \label{fig: fig5 72exp}
\end{figure}

Our experimental results follow similarly to simulation results; SBR-Net can reconstruct and localize emitters up to one scattering length. SBR-Net reconstructs emitters with SBR as low as 1.05 in experimental measurements of scattering phantoms, and as deep as one complete scattering length, as shown in Fig.~\ref{fig: fig4 279exp}b. Across all three scattering phantoms, SBR-Net can reconstruct emitters in 3D with higher fidelity as verified by the confocal measurements. Both FS-Net and the model-based algorithm are limited to reconstructing emitters in shallow layers and suffer from low sensitivity to low-SBR emitters likely because they have been removed by the background removal step. In addition, SBR-Net consistently recovers emitters with $25\mu$m axial resolution in all scattering phantoms, whereas the model-based algorithm suffers from much worse axial elongation ($>200 \mu$m) similar to our previous results\cite{xue_single-shot_2020}. For the $\ell_s = 72\mu$m phantom, since scattering length is much shorter than the axial elongation achieved by the model-based reconstruction, the result by this traditional method does not provide any meaningful axial information. In contrast, the much improved axial resolution by SBR-Net still enables 3D localization of emitters within the $72\mu$m scattering length range, as highlighted in Fig.~\ref{fig: fig5 72exp}c.

While SBR-Net localizes emitters well, it may reconstruct them with some inaccuracies in intensity and size. SBR-Net may also fail at having consistent accurate depth localization relative to neighboring emitters. Fig. \ref{fig: fig4 279exp}c shows an emitter encircled in yellow that is displaced axially relative to the correctly predicted group of emitters to its left and right. This can be explained by the emitter being one receptive field (RF) length ($344.45\mu$m) away from the edge of the volume. In addition, there are also refraction effects from the dome geometry of the phantom (see the XZ view of the confocal microscopy image of the whole sample in Fig. S7), which we do not account for in our forward model. Refraction causes the experimental measurement to have laterally displaced emitter locations compared to our free space model, which is more severe for the non-central microlenses’ view. Consequently, emitters near the edge, positioned beneath a surface with a normal vector that makes a large angle with the optical axis would suffer from more refraction, resulting in poor axial localization in the reconstruction.

In addition to these controlled phantom experiments, we also conducted a pilot study to demonstrate SBR-Net’s generalization capability on complex biological samples. We applied the simulator-trained SBR-Net directly on a $75\mu$m thick fixed section of mouse brain containing fluorescently labelled neurons expressing green fluorescent protein (GFP), as detailed in Figs.~S9-S11. We validated the reconstruction and 3D localization performance of SBR-Net using a confocal microscopy measurement as reference. Comparisons between SBR-Net trained on different SBR ranges, FS-Net, and model-based reconstructions are also provided. 

\subsection{Effect of training data SBR on network robustness-accuracy tradeoff}

\begin{figure}[t!]
    \includegraphics[width=1.0\linewidth]{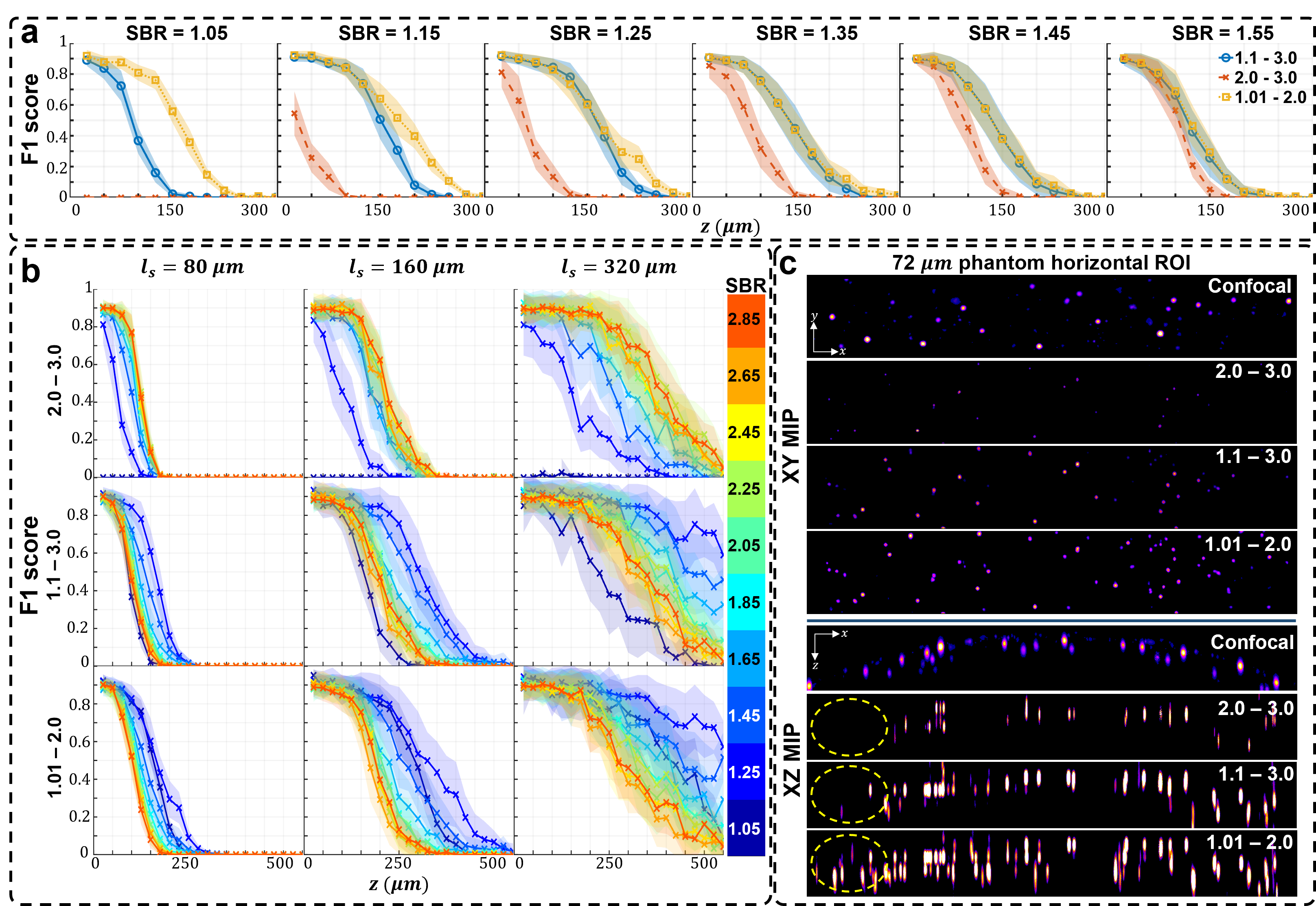}
  
  \caption{\textbf{Analysis of networks trained on different SBR ranges.} (a) The F1 score of each SBR-Net across different SBRs in simulation for $\ell_s=80\mu$m. SBR-Net trained on the lowest range of 1.01 -- 2.0 performs better than the other SBR-Nets. (b) On simulated testing data: F1 score of each SBR-Net (rows) for different scattering lengths (columns). SBR-Net (2.0 -- 3.0) has a depth penetration performance that is intuitive, performing better on data with higher SBR. In contrast, SBR-Net (1.01 -- 2.0) performs better on lower SBR than it does on higher SBR. SBR-Net (1.1 -- 3.0) has better depth penetration for data with SBR of 1.25 and 1.35 compared to data with higher SBR; this is similar to the behavior of SBR-Net (1.01 -- 2.0). However, for data with SBR 1.05 that is out of the training SBR range, it has the worse depth penetration performance compared to data with higher SBR. (c) On experimental testing data: XY and XZ MIPs of the reconstruction of the 72$\mu$m scattering phantom. SBR-Net (2.0 -- 3.0) behaves more conservatively, reconstructing emitters with high enough optical contrast and having fewer false positives at the expense of reconstructing lower SBR emitters. Examining the XY MIP of SBR-Net (1.01 -- 2.0) reconstruction, it appears to reconstruct all the emitters that are there in the confocal measurement, in addition to more false positives, highlighted in the dashed yellow oval. The XZ MIP displays a clearer visualization of this hallucination behavior. Our main result is trained on SBR between 1.1 and 3.0, which is a balance between the aforementioned two cases in experimental data.}
  \label{fig: fig6 sbr range}
\end{figure}

The choice of the SBR range on which to train SBR-Net was decided based on minimizing hallucination artifacts in the reconstructions of experimental data while also achieving good depth penetration. This is equivalent to balancing the robustness-accuracy tradeoff\cite{su_is_2019}, where, in our case, high robustness of the model is measured by the level of hallucination artifacts and accuracy measured by imaging depth penetration and emitter 3D localization. Hallucination artifacts are a sign of poor robustness because the model falsely predicts emitters from background fluorescence. This model has not learned a general enough background feature space to reject background or a general enough target emitter feature space; instead, it is overly sensitive to minute background fluctuations and attributes them as emitters in a non-consistent manner. We experimented with three SBR ranges based on typical SBRs seen in scattering phantom experiments and \textit{in vivo} rodent brain widefield imaging experiments. \textit{In vivo} one-photon imaging experiments usually present neurons with very weak signal contrast, having SBRs between 1.05 and 2\cite{greene_pupil_2023,kauvar_cortical_2020}, so we train one SBR-Net with a lower SBR range of 1.01 and 2.0. We also train an SBR-Net on a higher range between 2.0 -- 3.0, and between 1.1 -- 3.0, the latter being our main result.

While the SBR-Net (1.01 -- 2.0) performs better on synthetic data as shown in Fig. \ref{fig: fig6 sbr range}a-b, we observe many more false positives, or hallucinations, when it is applied to experimental data, as seen in Fig. \ref{fig: fig6 sbr range}c. In other words, SBR-Net (1.01 -- 2.0) has higher accuracy but lower robustness measure. In contrast, SBR-Net (2.0  -- 3.0) when applied to experimental data performs more conservatively at the cost of depth penetration, having higher robustness at the cost of accuracy. In our main results, we balance this trade-off using an SBR-Net (1.1 -- 3.0) trained on SBRs between 1.1 and 3.0. This network achieves depth penetration comparable to that in confocal microscopy with much fewer hallucination artifacts, as shown in Fig. \ref{fig: fig6 sbr range}c. An explanation for poor robustness of SBR-Net (1.01  -- 2.0) is due to the model being sensitive to small deviations in the real measurement from the synthetic training data, such as refraction due to non-planar sample surface, optics and sample induced aberrations, unseen SBR measurement, and sensor noise. Visually, the emitters are nearly imperceptible among background with SBR below 1.05. While this network performs well at depth penetration on synthetic test data that is generated from the same distribution as the training data, this sensitivity of emitter detection exacerbates the hallucination problem when the experimental data has even small distributional differences from the synthetic training data. We further carry out reconstruction of the three SBR-Nets on a 75$\mu$m thick brain slice sample and observe the same robustness-accuracy tradeoff, shown in Fig. S11.


\subsection{Number of unique training pairs affect bias-variance tradeoff}
It is well-known that the amount of training data has a major impact on neural network generalization\cite{lei_how_2022}. Therefore, we experiment with the number of unique training pairs to test the model's generalization performance. The goal is to balance the bias-variance tradeoff, where a good model has a high enough bias to produce low variance predictions, being more tolerant of OOD experimental data. We find that 500 unique training pairs allows good generalization to experimental data because there is not a diversity of features to learn, and we wish to avoid overfitting to synthetic training data as the network's behavior will become too brittle to generalize to experimental data, which has small distributional differences with synthetic data. 

We are mainly interested in learning features similar to spherical cellular bodies and background fluorescence, allowing the learning effort to require fewer examples compared to typical computer vision tasks that require large image datasets with more diverse features like Imagenet. We carry out an experiment where we change only the number of unique training data pairs; our main result is trained on 500 unique pairs (80/20, train/validate) and we train the same network with 1500 unique pairs (80/20, train/validate). We specify ``unique'' pairs because convolutional neural networks (CNNs) with max-pooling are equivariant in translation.  This means that even though 224 $\times$ 224 pixel patches from the entire 512 $\times$ 512 data are chosen randomly online during training, allowing a larger number of different examples for the network to learn from, smaller sub-patches are seen multiple times in different locations over different epochs, which do not contribute to an overall lower validation loss due to the translation equivariance of our CNN with 3 $\times$ 3 convolution kernels. Fig. S12 shows that SBR-Net trained on 1500 unique pairs suffers from hallucination artifacts, demonstrating high variance predictions and poor generalization to experimental data compared to our main result of SBR-Net trained on only 500 unique pairs. This is an indication that the bias of the model trained on 1500 pairs is lower, resulting in a less robust performance. From our analysis in previous sections, we understand that a lack of robustness results more hallucination artifacts. 

We can measure the level of overfitting to synthetic data by comparing the validation losses between the networks. In Fig. S14, we verify that the SBR-Net trained on 1500 data pairs has lower validation losses, i.e., fitting to synthetic data more. While this is desirable in most deep learning tasks, this validation set is also synthetic data. On the other hand, we desire learning the shared features between synthetic and experimental data without the network having a low enough bias where it specializes in synthetic data features, resulting in high variance inferences on experimental data. If more training pairs were used to train a model or if the model had more parameters, we may still control the bias-variance tradeoff by early stopping or other forms of regularization. However, our validation set can only be synthetic data, so this strategy is not practical due to the lack of ground truth for experimental data to decide when to stop early.

We focus on experiments with overfitting rather than underfitting because in practice, underfitting is unlikely to occur due to the lack of diversity of features in our learning task. While our choice of network architecture and number of unique training pairs may not be optimal, these results demonstrate the necessary considerations one should take for a model trained on synthetic data to be robust to experimental data.

\begin{figure}[t!]
    \centering
  \includegraphics[width=1\linewidth]{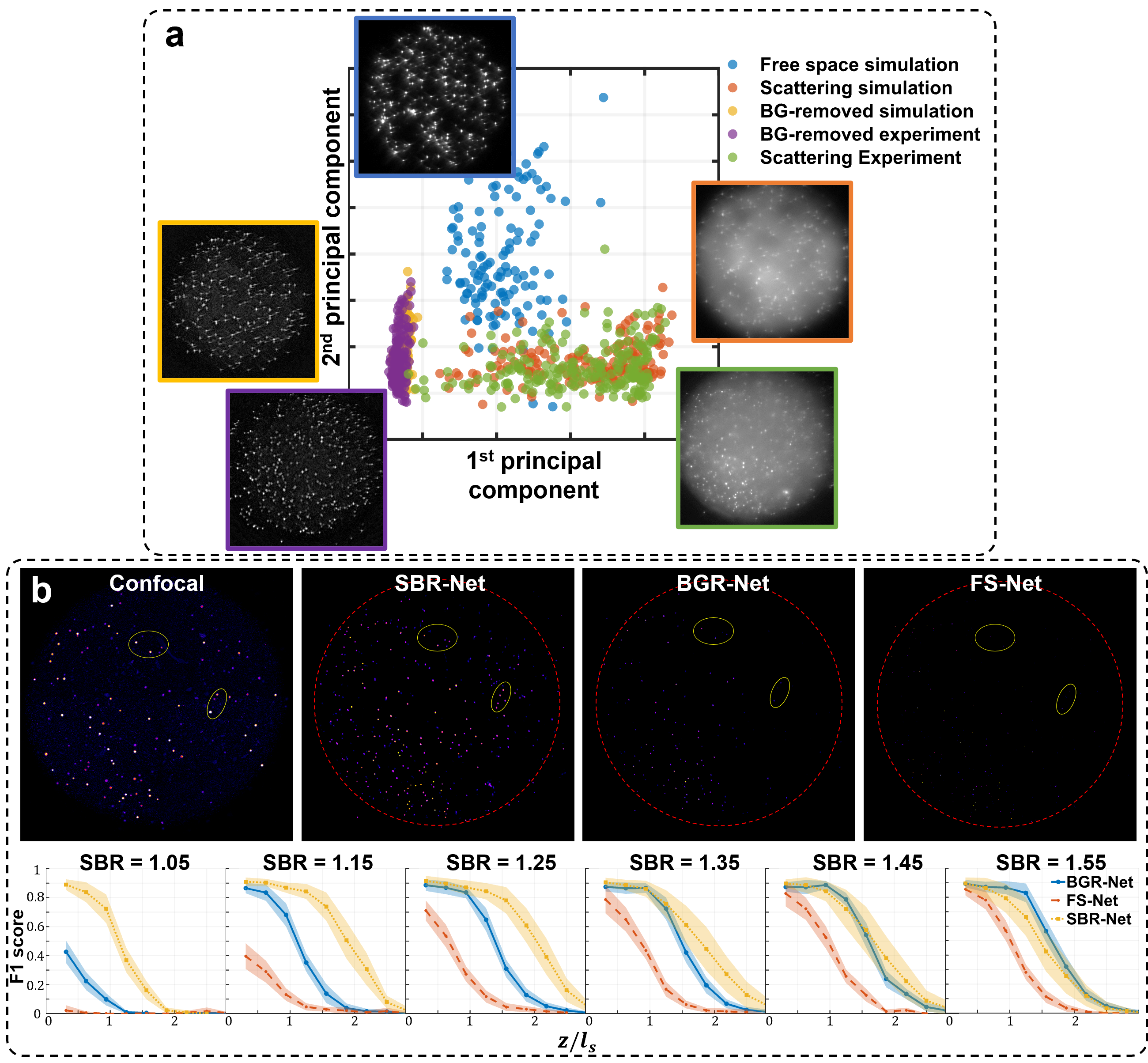}
  \caption{\textbf{Dimensional reduction and BGR-Net reconstruction comparison.} a) PCA of all measurement domains. b) Results of BGR-Net on the 72$\mu$m phantom. XY MIP reconstructions and F1 scores of BGR-Net are compared to SBR-Net and FS-Net. }
  \label{fig: pca and bgr test}
\end{figure}

\subsection{Input measurement analysis}
To gain a deeper understanding of the generalizability of our DL models, we perform dimensionality reduction through PCA for all measurements domains to analyze their similarities in an interpretable subspace. We compute the center location of all emitters in each measurement and retrieve a 32 $\times$ 32 pixel patch centered around it. Plotting each emitter's first and second principal component, we see similar clustering for the synthetic and experimental scattering measurements as shown in Fig. \ref{fig: pca and bgr test}a, providing another explanation for how our model trained on synthetic data generalizes to real experimental data.

One thing we observe is that the synthetic free-space data cluster is disjoint from the background removed data clusters. Background removed data is clearly an OOD input to FS-Net that was trained on synthetic free-space data, and our qualitative observations of false positives from FS-Net reconstructions of synthetic data in Fig. \ref{fig:fig2 qualitative} is consistent with experimental data where OOD inputs may lead to reconstructions with hallucinations.

Interestingly, we observe even more similar and tighter clustering between background-removed synthetic and experimental data. Naturally, we would want to explore if a network trained on scattering measurements that have undergone background removal would perform as well as SBR-Net. Thus, we train another background-removed network BGR-Net, and visualize its XY MIP reconstructions of the experimental $\ell_s = 72\mu$m phantom, using the confocal measurement as a reference baseline in Fig. \ref{fig: pca and bgr test}b. In simulation and experiment, BGR-Net is able to retain more emitters compared to FS-Net, but SBR-Net still performs better for reconstruction and localization. From the quantitative analysis of simulated data in Fig. \ref{fig: pca and bgr test}b, SBR-Net performs better than BGR-Net until a high enough SBR of around 1.55, where target emitters have more optical contrast and are less likely to be removed in the background removal step. However, for practical applications such as \textit{in vivo} neural imaging, many neurons exhibit measurement SBRs lower than 1.55, rendering the proposed SBR-Net a more suitable approach for such a task.

\section{Conclusion}
We demonstrate a DL-based single-shot 3D reconstruction algorithm, SBR-Net, to recover emitters embedded in scattering media with different scattering lengths. SBR-Net is based on learning from synthetic data that was generated with our scattering simulator that captures the heterogenous low-contrast signal and strong background behavior of fluorescence measurements in imaging through tissue scattering applications. The experimentally achieved depth penetration and optical sectioning performance of the simulator-trained SBR-Net is similar to that of confocal microscopy, while requiring only a single measurement to reconstruct the 3D volume. 

We analyze the generalization tradeoffs that arise from the choice of training data SBR range to discover that a larger range of SBRs between 1.1 and 3.0 lead to a more robust reconstruction for experimental data. Analyzing the effects of the number of training data pairs and number of model parameters reveals the well-known bias-variance tradeoff in machine learning. For our task of reconstructing spherical emitters and rejecting slowly varying background, we require fewer training data pairs and fewer model parameters since the diversity of learned features is limited and a more minimal learning structure is more suitable to generalize to experimental data and avoid overfitting to synthetic training data.

We conduct pilot study on the simulator-trained SBR-Net to biological samples. Results on the brain slice experiments show favorable progress towards neuronal imaging applications in rodent brains, where target signals have very low contrast, as shown in Figs.~S9 and S10, where SBR-Net performs significantly better than model-based reconstruction and FS-Net in both reconstruction and localization. While SBR-Net is able to recover cell bodies with strong 2D localization performance as shown in the XY MIP of the 3D reconstruction of Figs. S9 - S11, the severely low SBRs, compounded with uncontrolled experimental factors, such as random interfacial refraction, cell density, and unseen background features place the data even further OOD, resulting in sub-optimal 3D localization performance.

One caveat of our approach is that we only model the low-contrast behavior of scattered light. While this strategy has demonstrated favorable results in our phantom experiments, we believe that for improved performance across a broader range of applications, it is necessary to model optical aberrations caused by the imaging optics and the sample\cite{hampson_adaptive_2021}. Future work would investigate learning such aberrations based on physics-based simulator or experimental data, to incorporate them into a more robust model.

\begin{backmatter}
\bmsection{Funding}
National Institutes of Health (R01NS126596). J.A. acknowledges funding from the NSF Graduate Research Fellowship Program (GRFP) under Grant No. 2234657.

\bmsection{Acknowledgments}
The authors thank Boston University Shared Computing Cluster for proving computational resources. We also thank Anderson Chen for help with confocal microscopy.
Research reported in this publication was supported by the Boston University Micro and Nano Imaging Facility and the Office of the Director, National Institutes of Health of the National Institutes of Health under award Number S10OD024993. The content is solely the responsibility of the authors and does not necessarily represent the official views of the National Institute of Health.

\bmsection{Disclosures}
The authors declare no conflict of interest.

\bmsection{Ethics approval and consent to participate}
This study was performed in strict accordance with the recommendations in the Guide for the Care and Use of Laboratory Animals of the National Institutes of Health. All animals were handled according to approved Institutional Animal Care and Use Committee (IACUC) protocols (\#201800540) of Boston University.

\bmsection{Data availability statement}
Code, synthetic training data, and experimental data may be found at \href{https://github.com/bu-cisl/sbrnet}{https://github.com/bu-cisl/sbrnet}.

\bmsection{Supplemental document}
See Supplement 1 for supporting content.

\end{backmatter}

\bibliography{SBRNet_R0}

\begin{thebibliography}{10}
\newcommand{\enquote}[1]{``#1''}

\bibitem{mertz_strategies_2019}
J.~Mertz, \enquote{Strategies for volumetric imaging with a fluorescence microscope,} {\protect\JournalTitle{Optica}} \textbf{6}, 1261--1268 (2019).

\bibitem{weisenburger_guide_2018}
S.~Weisenburger and A.~Vaziri, \enquote{A {Guide} to {Emerging} {Technologies} for {Large}-{Scale} and {Whole}-{Brain} {Optical} {Imaging} of {Neuronal} {Activity},} {\protect\JournalTitle{Annual Review of Neuroscience}} \textbf{41}, 431--452 (2018).

\bibitem{cheng_development_2019}
X.~Cheng, Y.~Li, J.~Mertz, \emph{et~al.}, \enquote{Development of a beam propagation method to simulate the point spread function degradation in scattering media,} {\protect\JournalTitle{Optics letters}} \textbf{44}, 4989--4992 (2019).

\bibitem{horton_vivo_2013}
N.~G. Horton, K.~Wang, D.~Kobat, \emph{et~al.}, \enquote{In vivo three-photon microscopy of subcortical structures within an intact mouse brain,} {\protect\JournalTitle{Nature Photonics}} \textbf{7}, 205--209 (2013).

\bibitem{xue_single-shot_2020}
Y.~Xue, I.~G. Davison, D.~A. Boas, and L.~Tian, \enquote{Single-shot {3D} wide-field fluorescence imaging with a {Computational} {Miniature} {Mesoscope},} {\protect\JournalTitle{Science Advances}} \textbf{6}, eabb7508 (2020).

\bibitem{skocek_high-speed_2018}
O.~Skocek, T.~Nöbauer, L.~Weilguny, \emph{et~al.}, \enquote{High-speed volumetric imaging of neuronal activity in freely moving rodents,} {\protect\JournalTitle{Nature Methods}} \textbf{15}, 429--432 (2018).

\bibitem{kauvar_cortical_2020}
I.~V. Kauvar, T.~A. Machado, E.~Yuen, \emph{et~al.}, \enquote{Cortical {Observation} by {Synchronous} {Multifocal} {Optical} {Sampling} {Reveals} {Widespread} {Population} {Encoding} of {Actions},} {\protect\JournalTitle{Neuron}} \textbf{107}, 351--367.e19 (2020).

\bibitem{moretti_readout_2020}
C.~Moretti and S.~Gigan, \enquote{Readout of fluorescence functional signals through highly scattering tissue,} {\protect\JournalTitle{Nature Photonics}} \textbf{14}, 361--364 (2020).

\bibitem{li_deep-3d_2022}
B.~Li, S.~Tan, J.~Dong, \emph{et~al.}, \enquote{Deep-{3D} microscope: {3D} volumetric microscopy of thick scattering samples using a wide-field microscope and machine learning,} {\protect\JournalTitle{Biomedical Optics Express}} \textbf{13}, 284--299 (2022).

\bibitem{li_deep_2018}
Y.~Li, Y.~Xue, and L.~Tian, \enquote{Deep speckle correlation: a deep learning approach toward scalable imaging through scattering media,} {\protect\JournalTitle{Optica}} \textbf{5}, 1181--1190 (2018).

\bibitem{tahir_adaptive_2022}
W.~Tahir, H.~Wang, and L.~Tian, \enquote{Adaptive {3D} descattering with a dynamic synthesis network,} {\protect\JournalTitle{Light: Science \& Applications}} \textbf{11}, 42 (2022).

\bibitem{wijethilake_deep2_2022}
N.~Wijethilake, M.~Anandakumar, C.~Zheng, \emph{et~al.}, \enquote{{DEEP}{\textbackslash}ˆ2{\textbackslash}: {Deep} {Learning} {Powered} {De}-scattering with {Excitation} {Patterning},}  (2022). Issue: arXiv:2210.10892 arXiv: 2210.10892 [cs].

\bibitem{zhang_rapid_2023}
Y.~Zhang, G.~Zhang, X.~Han, \emph{et~al.}, \enquote{Rapid detection of neurons in widefield calcium imaging datasets after training with synthetic data,} {\protect\JournalTitle{Nature Methods}} pp. 1--8 (2023).

\bibitem{liu_recovery_2022}
R.~Liu, Y.~Sun, J.~Zhu, \emph{et~al.}, \enquote{Recovery of continuous {3D} refractive index maps from discrete intensity-only measurements using neural fields,} {\protect\JournalTitle{Nature Machine Intelligence}} \textbf{4}, 781--791 (2022).

\bibitem{xue_deep-learning-augmented_2022}
Y.~Xue, Q.~Yang, G.~Hu, \emph{et~al.}, \enquote{Deep-learning-augmented computational miniature mesoscope,} {\protect\JournalTitle{Optica}} \textbf{9}, 1009--1021 (2022).

\bibitem{yanny_miniscope3d_2020}
K.~Yanny, N.~Antipa, W.~Liberti, \emph{et~al.}, \enquote{{Miniscope3D}: optimized single-shot miniature {3D} fluorescence microscopy,} {\protect\JournalTitle{Light: Science \& Applications}} \textbf{9}, 171 (2020).

\bibitem{guo_fourier_2019}
C.~Guo, W.~Liu, X.~Hua, \emph{et~al.}, \enquote{Fourier light-field microscopy,} {\protect\JournalTitle{Optics Express}} \textbf{27}, 25573--25594 (2019).

\bibitem{nobauer_video_2017}
T.~Nöbauer, O.~Skocek, A.~J. Pernía-Andrade, \emph{et~al.}, \enquote{Video rate volumetric {Ca2}+ imaging across cortex using seeded iterative demixing ({SID}) microscopy,} {\protect\JournalTitle{Nature Methods}} \textbf{14}, 811--818 (2017).

\bibitem{pegard_compressive_2016}
N.~C. Pégard, H.-Y. Liu, N.~Antipa, \emph{et~al.}, \enquote{Compressive light-field microscopy for {3D} neural activity recording,} {\protect\JournalTitle{Optica}} \textbf{3}, 517--524 (2016).

\bibitem{zhang_computational_2021}
Y.~Zhang, Z.~Lu, J.~Wu, \emph{et~al.}, \enquote{Computational optical sectioning with an incoherent multiscale scattering model for light-field microscopy,} {\protect\JournalTitle{Nature Communications}} \textbf{12}, 6391 (2021).

\bibitem{noauthor_value_2021}
\enquote{Value noise,}  (2021).

\bibitem{noauthor_httpsgithubcombu-cislcomputational-miniature-mesoscope-cm2_nodate}
\enquote{https://github.com/bu-cisl/{Computational}-{Miniature}-{Mesoscope}-{CM2},} .

\bibitem{mockl_accurate_2020}
L.~Möckl, A.~R. Roy, P.~N. Petrov, and W.~E. Moerner, \enquote{Accurate and rapid background estimation in single-molecule localization microscopy using the deep neural network {BGnet},} {\protect\JournalTitle{Proceedings of the National Academy of Sciences}} \textbf{117}, 60--67 (2020).

\bibitem{262588213843476_2d_nodate}
{262588213843476}, \enquote{{2D} and {3D} {Perlin} {Noise} in {MATLAB},} .

\bibitem{foi_practical_2008}
A.~Foi, M.~Trimeche, V.~Katkovnik, and K.~Egiazarian, \enquote{Practical {Poissonian}-{Gaussian} noise modeling and fitting for single-image raw-data,} {\protect\JournalTitle{IEEE transactions on image processing: a publication of the IEEE Signal Processing Society}} \textbf{17}, 1737--1754 (2008).

\bibitem{he_delving_2015}
K.~He, X.~Zhang, S.~Ren, and J.~Sun, \enquote{Delving {Deep} into {Rectifiers}: {Surpassing} {Human}-{Level} {Performance} on {ImageNet} {Classification},}  (2015). Issue: arXiv:1502.01852 arXiv: 1502.01852 [cs].

\bibitem{loshchilov_sgdr_2017}
I.~Loshchilov and F.~Hutter, \enquote{{SGDR}: {Stochastic} {Gradient} {Descent} with {Warm} {Restarts},}  (2017). Issue: arXiv:1608.03983 arXiv: 1608.03983 [cs, math].

\bibitem{deng_learning_2020}
M.~Deng, S.~Li, A.~Goy, \emph{et~al.}, \enquote{Learning to synthesize: robust phase retrieval at low photon counts,} {\protect\JournalTitle{Light: Science \& Applications}} \textbf{9}, 36 (2020).

\bibitem{su_is_2019}
D.~Su, H.~Zhang, H.~Chen, \emph{et~al.}, \enquote{Is {Robustness} the {Cost} of {Accuracy}? – {A} {Comprehensive} {Study} on the {Robustness} of 18 {Deep} {Image} {Classification} {Models},}  (2019). Issue: arXiv:1808.01688 arXiv: 1808.01688 [cs].

\bibitem{greene_pupil_2023}
J.~Greene, Y.~Xue, J.~Alido, \emph{et~al.}, \enquote{Pupil engineering for extended depth-of-field imaging in a fluorescence miniscope,} {\protect\JournalTitle{Neurophotonics}} \textbf{10}, 044302 (2023). Publisher: SPIE.

\bibitem{lei_how_2022}
S.~Lei, H.~Zhang, K.~Wang, and Z.~Su, \enquote{How {Training} {Data} {Affect} the {Accuracy} and {Robustness} of {Neural} {Networks} for {Image} {Classification},} {\protect\JournalTitle{""}}  (2022).

\bibitem{hampson_adaptive_2021}
K.~M. Hampson, R.~Turcotte, D.~T. Miller, \emph{et~al.}, \enquote{Adaptive optics for high-resolution imaging,} {\protect\JournalTitle{Nature Reviews Methods Primers}} \textbf{1}, 1--26 (2021).

\end{thebibliography}

\end{document}